# Parametric and nonparametric models and methods in financial econometrics[*,†]

Zhibiao Zhao

*Department of Statistics*
*Pennsylvania State University*
*e-mail:* zuz13@stat.psu.edu

**Abstract:** Financial econometrics has become an increasingly popular research field. In this paper we review a few parametric and nonparametric models and methods used in this area. After introducing several widely used continuous-time and discrete-time models, we study in detail dependence structures of discrete samples, including Markovian property, hidden Markovian structure, contaminated observations, and random samples. We then discuss several popular parametric and nonparametric estimation methods. To avoid model mis-specification, model validation plays a key role in financial modeling. We discuss several model validation techniques, including pseudo-likelihood ratio test, nonparametric curve regression based test, residuals based test, generalized likelihood ratio test, simultaneous confidence band construction, and density based test. Finally, we briefly touch on tools for studying large sample properties.

Received January 2008.

## Contents

---

[*]This is an original survey paper.
[†]This paper was accepted by Donald Richards, Associate Editor for the IMS.

## 1. Introduction

Over the past few decades, financial econometrics has become an increasingly popular research field among the economics, finance, statistics and probability communities, and such a trend will undoubtedly continue. We refer to the books by Franke *et al.* [84], Hull [105] and Neftci [138] for an elementary introduction to mathematical finance; Duffie [60] for asset pricing theory; Steele [150], Karatzas and Shreve [116] and Karlin and Taylor [117] for extensive treatments of stochastic calculus and martingales; Fan and Yao [77] and Li and Racine [127] for nonparametric methods in time series; Gao [87] for semi-parametric methods in econometrics; and Tsay [155] for an excellent exposition of financial time series analysis among others. In this survey paper, we provide a selective overview of some popular parametric and nonparametric models and methods in financial econometrics.

One of the main objectives of financial econometrics is to understand and model the evolving dynamics behind the financial markets. To model the price dynamics of assets that are subject to uncertainties, various continuous-time models in the form of stochastic differential equations and discrete-time series models have been proposed with the hope that they could provide a reasonable approximation to the true data-generating dynamics. In Sections 2 and 3, we review some popular continuous-time and discrete-time models, respectively. In practice, since only discrete observations are available, dependence structures of such discrete observations are discussed in Section 4.

Given a model, a natural problem is to estimate the unknown quantities of the model based on discrete observations. Section 5 reviews various parameter estimation methods when we have sufficient prior knowledge that the model has a parametric form with unknown parameters. Nonparametric models can reduce modeling bias by imposing no specific model structure other than certain smoothness assumptions, and therefore they are particularly useful when we have little information or we want to be reflexible about the underlying model. Section 6 gives a brief account of some useful nonparametric methods. Since the payoff for derivatives depends critically on the price process of the underlying security, it is very important that the price model of the underlying security be correctly specified. Such model validation problems are addressed in Section 7. Section 8 contains some useful tools for the study of large sample properties of parametric and nonparametric estimates.

Fan [75] gives an excellent overview of nonparametric methods in financial econometrics. The present paper adds new material that was not covered by Fan [75], including jump diffusion models, stochastic volatility models, discrete-time models, dependence structure of discrete samples, more detailed discussion on parametric and nonparametric methods and model validations, and some tools for studying large sample properties.

## 2. Continuous-time models

### 2.1. Continuous-time diffusion models

A European call option gives the holder the right to buy the underlying asset $S_t$, at the expiration date, or maturity, $T$ for a certain strike price $K$, but the holder does not have to exercise that right. Therefore, the payoff for a European option written on $S_t$ is $\max(S_T - K, 0)$, and this payoff depends critically on the behavior of the underlying asset $S_t$. For an introduction to financial derivatives, see Hull [105].

As a milestone of quantitative finance, Black and Scholes [39] assume the following model for $S_t$ to derive their celebrated pricing formula for European call options:

$$dS_t = \mu S_t dt + \sigma S_t dW_t, \tag{2.1}$$

where $\{W_t\}_{t\geq 0}$ is a standard Brownian motion, and $\mu$ and $\sigma$ are the drift and diffusion coefficients, respectively. By Itô's Lemma, the solution of (2.1) is the geometric Brownian motion (GBM)

$$S_t = \exp[(\mu - \sigma^2/2)t + \sigma W_t]. \tag{2.2}$$

As the simplest model for modeling stock prices, GBM is still widely used in the modern financial community.

Under the "risk-neutral" world, we would expect that the interest discounted payoff with the discount factor $\exp(-\int_0^T r_t dt)$ for compound interest $r_t$ should be a martingale and hence all stocks earn the same rate as the risk-free rates $r_t$. In the early literature of finance, interest rates are considered to be constant; see [39] and [135]. The latter assumption works reasonably well if we consider only a short period of time during which interest rates remain approximately the same, while it causes important discrepancies over a long time span. For example, interest rates on one-year U.S. Treasury bills ranged from as high as 15% in the early 1980's to as low as less than 1% in 2003. Interest rates are not tradable assets, while derivatives (for example, interest rate swaps, futures, bond options) written on them are. In fact, the interest rate derivatives market is the largest derivatives market with an average daily turnover of about $ 60 trillion dollars; see [81].

To price interest rate derivatives, Vasicek [156] proposes the following model for interest rates $r_t$,

$$dr_t = \beta(\alpha - r_t)dt + \sigma dW_t, \quad \alpha > 0, \quad \beta > 0. \tag{2.3}$$

The quantity $\alpha$ determines the long-run average interest rate. If $r_t > \alpha$, then $\beta(\alpha - r_t) < 0$ pulls interest rates downward; while when $r_t < \alpha$, $\beta(\alpha - r_t) > 0$ pushes rates upward. Therefore, (2.3) bears a mean-reversion explanation with $\alpha$ being the mean value and $\beta$ the strength of the mean-reversion. Model (2.3) is the well-known Ornstein-Uhlenbeck process whose solution, assuming $X_0$ is Gaussian distributed with mean $\alpha$ and variance $\sigma^2/(2\beta)$, is a stationary Gaussian process fully characterized by the mean $\alpha$ and covariance function

$$\mathrm{Cov}(r_s, r_t) = \frac{\sigma^2}{2\beta} e^{-\beta|t-s|}.$$

Model (2.3) assumes constant volatility $\sigma$, which hardly matches empirical observations. For example, volatilities tend to be clustered and larger observations are associated with larger volatilities. A more realistic model would take into account the non-constant volatility. Cox, Ingersoll and Ross [53] derive the "CIR" model

$$dr_t = \beta(\alpha - r_t)dt + \sigma r_t^{1/2} dW_t. \tag{2.4}$$

Assume that $2\alpha\beta \geq \sigma^2$, (2.4) admits a non-negative solution $r_t$ that possess a noncentral chi-square transition density, while the marginal density is a Gamma density. Another model for interest rates used in Courtadon [52] is $dr_t = \beta(\alpha - r_t)dt + \sigma r_t dW_t$. Chan, Karolyi, Longstaff and Sanders [48] further extend the "CIR" model to the "CKLS" model,

$$dr_t = \beta(\alpha - r_t) + \sigma r_t^\gamma dW_t. \tag{2.5}$$

For other specifications of interest rate models, see [45; 51; 62; 133]. Aït-Sahalia [2] finds that the 7-day Euro-dollar deposit rate has a strong nonlinear mean-reversion only when the rate is beyond the range 4%–17%. To address this issue, he proposes the nonlinear drift model

$$dr_t = (\alpha_0 + \alpha_1 r_t + \alpha_2 r_t^2 + \alpha_3/r_t)dt + \sqrt{\beta_0 + \beta_1 r_t + \beta_2 r_t^\gamma}dW_t, \quad (2.6)$$

which includes all the aforementioned models as special cases.

A nonparametric one-factor diffusion model has the form

$$dX_t = \mu(X_t)dt + \sigma(X_t)dW_t. \quad (2.7)$$

Here $\{X_t\}$ might be stock prices, interest rates, the S&P 500 index, or other financial quantities; $\mu(\cdot)$ and $\sigma(\cdot)$ are termed drift or instantaneous return function, and diffusion, volatility, or instantaneous return variance function, respectively. In particular, a parametric counterpart of (2.7) is

$$dX_t = \mu_\theta(X_t)dt + \sigma_\theta(X_t)dW_t, \quad (2.8)$$

where $(\mu_\theta, \sigma_\theta)$ is a known parametric specification with unknown parameter $\theta \in \mathbb{R}^d$. By proper specifications of $(\mu_\theta, \sigma_\theta)$, we can recover all the parametric models discussed above. Conditions for existence of weak solutions and strong solutions have been derived in Karlin and Taylor [117]. If the price process is assumed to follow the diffusion model (2.7), then among the key quantities of interest are the unknown drift $\mu(\cdot)$, diffusion $\sigma(\cdot)$, and the probabilistic properties of $X_t$. For option pricing, it is very important that the price model of the underlying security be correctly specified parametrically (model validation) or consistently estimated nonparametrically (nonparametric estimation). Such model estimation and validation questions are addressed in Sections 5, 6 and 7.

## 2.2. Jump diffusion processes

There was little doubt that the stock returns based on logarithm are independent Gaussian random variables until the year of 1963 when Mandelbrot published his classical paper [132]. Mandelbrot [132] studied the cotton price changes and found that: (i) The histograms of price changes are too peaked relative to Gaussian distributions; and (ii) the tails of the distributions of the cotton price changes are so extraordinarily long that it may be reasonable to assume that the second moment is infinite. Mandelbrot [132] further argued that a good alternative model for cotton price changes is the stable distribution with index 1.7, pioneering the approach of modeling financial data using Lévy processes. See also the discussion in Fama [73].

A Lévy process is a stochastic process with right-continuous sample path and independent and stationary increments. Special examples of Lévy processes include Brownian motions, Poisson processes, and stable processes among others. The latter two processes are jump type processes. Here we focus on diffusion processes with Poisson jumps. Such jump diffusion processes have been proposed to capture the heavy-tailedness feature of returns. Other types of Lévy processes based models have also been proposed. See, for example, Eberlein and Keller [66] and Eberlein *et al.* [67] for hyperbolic Lévy motions and their applications in fitting German stock returns; Aït-Sahalia and Jacod [7; 8], Eberlein *et al.* [65], Eberlein and Raible [68], Nolan [140], Woerner [159] for estimations under Lévy process settings. Zhao and Wu [174] study nonparametric inferences for nonstationary process driven by $\alpha$-stable Lévy processes; see also the references thereof.

The first attempt to incorporate jumps into diffusion model was made by Merton [136]. The basic idea is that we may assume there are two types of randomness driving the stock prices: the first is a Brownian motion generating continuous sample path and small movements, while the second one is large but infrequent jumps representing sudden shocks/news. In particular, [136] assumes that stock prices follow the following jump diffusion model

$$dS_t/S_{t-} = (\alpha - \lambda\kappa)dt + \sigma dW_t + J_t dN_t, \qquad (2.9)$$

where $N_t$ is a counting Poisson process for jumps with intensity $\lambda$, $J_t$ is an independent jump size if a jump occurs at $t$, the jumps are assumed to be iid, and $\kappa = \mathbb{E}(J_t)$. The inclusion of the coefficient $\lambda\kappa$ in the drift makes $S_t$ unpredictable.

By the Doléans-Dade formula, (2.9) admits the following solution

$$S_t/S_0 = \exp[(\alpha - \sigma^2/2 - \lambda\kappa)t + \sigma W_t] \prod_{i=1}^{N_t} J_{T_i},$$

with the convention $\prod_{i=1}^{0} = 1$, where $T_i, 1 \leq i \leq N_t$, denote the times at which jumps occur. A typical choice of $J_t$ is lognormal random variable so that $S_t/S_0$ has lognormal distribution. Merton [136] derives an option pricing formula for call options written on a security whose price process follows (2.9).

Over the past three decades, (2.9) has been extended in various directions by specifying different structures for the drift, diffusion, and jump components. For example, assuming that the magnitudes of the jumps are dependent, Oldfield *et al.* [141] propose an autoregressive jump diffusion model. Ball and Torous [22] replace the Poisson jump process by a Bernoulli jump process and argue that the latter process could yield more satisfactory empirical and theoretical analysis, including computational advantages and the attainment of the Cramér-Rao lower bound for maximum likelihood estimation. Ramezani and Zeng [143] and Kou [122] use an asymmetric double exponential distribution for $\log(J_t)$, and show that the resulting model can capture asymmetric leptokurtic features and "volatility smile" features frequently observed in financial data. In Bates [32], the volatility $\sigma$ in (2.9) is assumed to be a stochastic process of a mean-reverting type,

$$d\sigma_t^2 = (\theta - \beta\sigma_t^2)dt + s\sqrt{\sigma_t^2}dW_t', \qquad (2.10)$$

for another Brownian motion $\{W_t'\}$ which could be correlated with $\{W_t\}$. Jorion [115] performs a significance test of jump components and concludes that many exchange rates display significant jump components; see also Lee and Mykland [124]. Duffie *et al.* [64] study option pricing for multi-dimensional affine jump diffusion models in which the drift vector, volatility matrix and jump components are assumed to be affine functions of the state variable $S_t$. For other contributions on jump diffusion processes and their applications, see [14; 21; 31; 33; 110; 111; 129] and references therein.

## 2.3. Continuous-time stochastic volatility models

Stochastic volatility (SV) model has emerged as a powerful alternative to the traditional deterministic volatility model. In contrast to deterministic volatility models that assume the volatility is a deterministic function of the stock prices, SV models assume that the volatility is also a stochastic process. The paper by Hull and White [106] is among the first to study SV models. They consider the model

$$dS_t = \mu S_t dt + \sigma_t S_t dW_t \quad \text{and} \quad d\sigma_t^2 = \beta \sigma_t^2 dt + \nu \sigma_t^2 dW_t', \qquad (2.11)$$

where $\{W_t\}$ and $\{W_t'\}$ are two standard Brownian motions whose increments have correlation $\rho$. The volatility $\{\sigma_t^2\}$ is a geometric Brownian motion. Hull and White [106] find that the pricing formula for a European call option under the SV model (2.11) behaves differently from the classical Black-Scholes (B-S) formula. Under B-S formula, at-the-money options tend to be overpriced while deep-in-the-money and deep-out-of-the-money options tend to be underpriced.

Since Hull and White [106], various SV models have been proposed. For example, Scott [146] introduces a mean-reverting Ornstein-Uhlenbeck process for the volatility $d\sigma_t = \beta(\alpha - \sigma_t)dt + \nu dW_t'$, Melino and Turnbull [134] assume the CKLS type model [cf. (2.5)] $dS_t = \beta(\alpha - S_t)dt + \sigma_t S_t^\gamma dW_t$ for the Canada Dollar/U.S. Dollar spot exchange rates $S_t$ with the volatility process $\log(\sigma_t)$ being an Ornstein-Uhlenbeck process; see also Wiggins [158], and Andersen and Lund [16].

In high-frequency setting, instantaneous returns are usually negligible relative to volatilities and hence can be taken to be zero. Under this setting, a general nonparametric continuous-time diffusion model with stochastic volatility is

$$d\log(S_t) = \sigma_t dW_t, \quad d\log(\sigma_t^2) = r(\log(\sigma_t^2))dt + s(\log(\sigma_t^2))dW_t', \qquad (2.12)$$

where $\{W_t\}$ and $\{W_t'\}$ are two standard Brownian motions with correlation $\rho = \text{Corr}(dW_t, dW_t')$. When $\rho < 0$, this model is often used to model leverage effect. When bad news releases, equity price $S_t$ drops and $dW_t < 0$. The negative correlation then implies that $dW_t' > 0$ and hence an increase in volatility $\sigma_t^2$. For example, Yu [165] proposes $d\log(\sigma_t^2) = \alpha + \beta \log(\sigma_t^2)dt + \sigma_v dW_t'$, and Omori *et al.* [142] study its discrete version.

The aforementioned SV models are built out of Brownian motions and a natural extension is to replace Brownian motions by more general processes.

Barndorff-Nielsen and Shephard [28] introduce another class of SV models based on Lévy processes by assuming that the logarithm of an asset price follows

$$d\log(S_t) = (\mu + \beta\sigma_t^2)dt + \sigma_t dW_t \quad \text{and} \quad d\sigma_t^2 = -\lambda\sigma_t^2 dt + dZ_{\lambda t}, \qquad (2.13)$$

where $\lambda > 0$ and $\{Z_t\}$ is a general Lévy process with stationary and independent increments. The Lévy process $\{Z_t\}$ other than Brownian motions has jumps and therefore the volatility $\{\sigma_t^2\}$ may exhibit big jumps. On the other hand, due to the continuity of $\{W_t\}$, the asset price $\{S_t\}$ is still continuous. Barndorff-Nielsen and Shephard [28] show that many specific SV models can be built out of (2.13) by specifying either the marginal process $\{\sigma_t\}$ or the Lévy process $\{Z_t\}$.

In (2.12), let $X_i = \log(S_{i\Delta}) - \log(S_{(i-1)\Delta})$ be the aggregated log returns during time period $[(i-1)\Delta, i\Delta]$. The unobserved stochastic volatility process $\{\sigma_t^2\}_{t\geq 0}$ is a stationary Markov process. However, the returns $\{X_i\}$ do not form a Markov chain as in the deterministic volatility model (2.7). Instead, when $\{W_t\}$ and $\{W_t'\}$ are independent, they form a hidden Markov model; see Section 4.4. Finally, we point out that it is a common practice to use a discrete version of (2.12) to facilitate computational and theoretical derivations; see Section 3.2.

## 3. Discrete-time models

So far we have discussed continuous-time models. Another powerful tool in studying dynamics of variables in financial markets is time series analysis. In practice, all continuous-time models are observed at discrete times, therefore we may model these discretely observed measurements using time series models. In fact, despite the fact that (2.7) is written in a continuous-time form, one often uses the following Euler discretization scheme

$$X_{t+\Delta} - X_t = \mu(X_t)\Delta + \sigma(X_t)(W_{t+\Delta} - W_t), \quad t = 0, \Delta, 2\Delta, \ldots,$$

as an approximation to facilitate computational and theoretical derivation. The accuracy of such Euler discretization is studied in Jacod and Protter [107]. We devote this section to reviewing discrete time series models.

### 3.1. Nonlinear autoregressive and stochastic regression models

A discrete version of the continuous-time model (2.7) is the nonparametric autoregressive conditional heteroscedastic (NARCH) model

$$X_i = \mu(X_{i-1}) + \sigma(X_{i-1})\varepsilon_i, \qquad (3.1)$$

where $\varepsilon_i, i \in \mathbb{Z}$, are iid random variables. If $\sigma(\cdot)$ is a constant function, then (3.1) is called a nonparametric autoregressive (NAR) model. Special cases of (3.1) include linear AR model: $X_i = aX_{i-1} + \varepsilon_i$, threshold AR [Tong [154]] model: $X_i = a\max(X_i, 0) + b\min(X_i, 0) + \varepsilon_i$, and exponential AR [Haggan and Ozaki [93]] model: $X_i = [a + b\exp(-cX_{i-1})]X_{i-1} + \varepsilon_i$ among others. The latter three models have constant conditional variances. In an attempt to model

United Kingdom inflation during the time period 1958–1977, Engle [69] proposes the class of autoregressive conditional heteroscedastic (ARCH) models. The essential idea of ARCH models is that the conditional variances are non-constant but rather change as time evolves. In particular, ARCH model of order one has the form

$$X_i = \sqrt{\alpha_0^2 + \alpha_1^2 X_{i-1}^2} \varepsilon_i, \quad \alpha_0 > 0, 0 < \alpha_1 < 1. \tag{3.2}$$

Model (3.1) can generate heavy-tailed distributions. To see this, consider the simple model $X_i = \sigma(X_{i-1})\varepsilon_i$, and assume that $\varepsilon_0$ has standard normal distribution. Then, by Jensen's inequality,

$$\text{kurtosis}(X_i) = \frac{\mathbb{E}(X_i^4)}{[\mathbb{E}(X_i^2)]^2} = \frac{\mathbb{E}[\sigma^4(X_{i-1})\varepsilon_i^4]}{[\mathbb{E}(\sigma^2(X_{i-1})\varepsilon_i^2)]^2} = 3\frac{\mathbb{E}[\sigma^4(X_{i-1})]}{[\mathbb{E}(\sigma^2(X_{i-1}))^2]} \geq 3.$$

For example, consider the ARCH(1) model in (3.2). It is easy to show that $\mathbb{E}(X_i^4)/[\mathbb{E}(X_i^2)]^2 = 3(1-\alpha_1^2)/(1-3\alpha_1^2) > 3$ if $3\alpha_1^2 < 1$ and $\mathbb{E}(X_i^4)/[\mathbb{E}(X_i^2)]^2 = \infty$ if $3\alpha_1^2 \geq 1$. The heavy-tailedness feature implied by model (3.1) makes it a successful candidate in many financial applications where it is frequently observed that returns exhibit heavy tails; see [155].

A more general version of (3.1) is the following stochastic regression model of order one

$$X_i = \mu(Y_i) + \sigma(Y_i)\varepsilon_i. \tag{3.3}$$

Here $Y_i$ and $X_i$ are the covariate variable and response, respectively, and the error $\varepsilon_i$ is independent of $Y_j, j \leq i$. In the special case of $Y_i = X_{i-1}$, (3.3) reduces to (3.1). Depending on the context, we may model $(Y_i)_{i \in \mathbb{N}}$ as a sequence of either iid random variables or time series. For example, if $Y_i$ is the measurement for $i$-th subject, then we may assume that $(Y_i)_{i \in \mathbb{N}}$ are iid. On the other hand, if $Y_i$ is the measurement for a subject at time $i$, then it is natural to assume that $(Y_i)_{i \in \mathbb{N}}$ form a time series. In the latter case, a possible model for $Y_i$ might be the NARCH model

$$Y_i = \tilde{\mu}(Y_{i-1}) + \tilde{\sigma}(Y_{i-1})\eta_i, \tag{3.4}$$

where $\eta_i, i \in \mathbb{Z}$ are iid random variables. To ensure that $\varepsilon_i$ is independent of $Y_j, j \leq i$, we assume that $\varepsilon_i$ is independent of $\eta_j, j \leq i$. See Zhao and Wu [172; 173] and references therein.

Due to the so called "*curse of dimensionality*", it is practically infeasible to extend the nonparametric model (3.1) to orders beyond two. Other extensions of model (3.1) include ARCH model with order $p$, generalized ARCH (GARCH) model in Bollerslev[40] and exponential GARCH (EGARCH) model in Nelson [139] to new a few. For extensive expositions of applications of ARCH, GARCH models and their variants in financial econometrics, see the survey papers [34; 41; 42; 57; 148] and the books by Gourieroux [92] and Tsay [155] (Chapter 3). Semi-parametric approach is studied in Gao [87].

### 3.2. Discrete-time stochastic volatility models

An Euler discretization of (2.12) is the following discrete-time stochastic volatility model

$$X_i = \sigma_i \eta_i \quad \text{and} \quad \sigma_i^2 = r(\sigma_{i-1}^2) + s(\sigma_{i-1}^2)\varepsilon_i, \tag{3.5}$$

where $\{\varepsilon_i\}_{i\in\mathbb{Z}}$ and $\{\eta_i\}_{i\in\mathbb{Z}}$ are two iid sequences. To guarantee positivity of the volatility $\sigma_i^2$, it is common to model the logarithm of volatility instead. For example, Taylor [152] proposes an AR(1) model for $\log(\sigma_i^2)$:

$$X_i = \sigma_i \eta_i \quad \text{and} \quad \log(\sigma_i^2) = \lambda \log(\sigma_{i-1}^2) + \varepsilon_i. \tag{3.6}$$

Note that a constant term in the right hand side of (3.6) is unnecessary since such a term can always be absorbed into $\eta_i$ in (3.5). In (3.6), the innovations $\varepsilon_i, i \in \mathbb{Z}$, are assumed to be iid normals with mean zero and variance $\sigma_\varepsilon^2$ and independent of $\eta_i$ in (3.5). The variance $\sigma_\varepsilon^2$ of $\varepsilon_i$ measures the uncertainty of future volatility. In the special of $\sigma_\varepsilon^2 = 0$ and $\lambda = 1$, the volatility is a deterministic constant. Thanks to the linear autoregressive relationship, (3.6) is often called ARSV(1) model or lognormal stochastic autoregressive volatility (SARV) model. Due to its simple structure and mathematical tractability, (3.6) has been extensively studied in the literature; see Broto and Ruiz [46], Shephard [148], Taylor [153], and references thereof. Ball and Torous [22] incorporate the discrete version of Chan *et al.* [48] and propose the following SV model for interest rates:

$$r_t = (\alpha + \beta)r_{t-1} + \sigma_{t-1}r_{t-1}^\gamma \varepsilon_t \quad \text{and} \quad \log \sigma_t = \rho \sigma_{t-1} + \nu(1-\rho) + \eta_t. \tag{3.7}$$

Wiggins [158] also uses a similar model under continuous-time setting.

In (3.6), a common choice for the density of $\varepsilon_i$ is the standard normal density $\phi$. Then the conditional distribution of $X_i$ given $\sigma_i$ is Gaussian. Therefore, the marginal density, denoted by $f_X$, of $X_i$ is a mixture of normal densities $\sigma_i^{-1}\phi(x/\sigma_i)$ with respect to the density $f_\sigma$ of $\sigma_i$: $f_X(x) = \int \sigma^{-1}\phi(x/\sigma)f_\sigma(\sigma)d\sigma$. As in the case of ARCH model in Section 3.1, $\{X_i\}$ from (3.6) exhibit heavier tails than that of the normal errors $\{\varepsilon_i\}$. Other heavy-tailed distributions of the errors $\{\varepsilon_i\}$ are studied in, for example, Harvey *et al.* [100], Bardnorff-Nielsen [27], Gallant *et al.* [85], and Liesenfeld and Jung [128].

Recently, model (3.6) with $(\eta_i, \varepsilon_{i+1}), i \in \mathbb{Z}$, forming iid copies of bivariate normal vector $(\eta, \varepsilon)$, $\text{Cov}(\eta, \varepsilon) < 0$, have been introduced to model leverage effect; see Yu [165] and Omori *et al.* [142]. For applications of SV models and their estimations see the survey papers by Broto and Ruiz [46], Ghysels *et al.* [91], and Shephard [148].

### 4. Dependence structure of discrete samples

Let $\{X_t\}_{t\in\mathcal{T}}$ be a generic process with time index $\mathcal{T}$. For example, $\{X_t\}_{t\in\mathcal{T}}$ might be the continuous-time models in Section 2 with $\mathcal{T} = [0, \infty)$ or the discrete-time models in Section 3 with $\mathcal{T} = \{0, 1, 2, \ldots\}$. Regardless of whether $\mathcal{T}$ is continuous-time or discrete-time, in practice, the process $\{X_t\}_{t\in\mathcal{T}}$ is only observed at discrete time points which could be possibly random. Moreover, the observations could be contaminated with errors. One of the main objectives in financial modeling is to study the dependence structure behind the data-generating mechanism based on discrete observations. In this section we review some main dependence features for discrete observations from financial models.

## *4.1. Markov chains*

Markov chains are widely used in virtually every scientific subjects, including biology, engineering, queueing theory, physics among others. In financial econometrics, due to the property of independent and stationary increments of Brownian motions, it would be reasonable to expect that the Markovian property is in the cards. In fact, we deserve even more: under some growth conditions on the drift $\mu$ and diffusion $\sigma$, $\{X_t\}_{t\geq 0}$ defined by the stochastic differentiation equation (2.7) is strong Markovian [for a definition, see pp. 149–152 in Karlin and Taylor [117]]. Therefore, discrete observations $\{X_i \equiv X_{i\Delta}\}_{i\geq 0}$ form a Markov chain. Here $\Delta > 0$ is a small but fixed number representing sampling frequency. The process $\{X_i\}_{i\geq 0}$ in (3.1) is also Markovian.

The Markovian property plays an important role in statistical estimation and inference. Let $\{X_i\}_{i\in\mathbb{Z}}$ be a stationary Markov chain. Denote by $\pi(x;\theta)$ and $p(x|x';\theta)$ the marginal density function of $X_0$ and the transition density function of $X_{i+1}$ at $x$ given $X_i = x'$, respectively. Here $\theta$ is parameter. Given observations $X_i, 0 \leq i \leq n$, the log likelihood function is given by

$$\begin{aligned}\ell(X_0,\ldots,X_n;\theta) &= \sum_{i=1}^{n}\log[p(X_i|X_{i-1};\theta)] + \log[\pi(X_0;\theta)]\\ &\approx \sum_{i=1}^{n}\log[p(X_i|X_{i-1};\theta)].\end{aligned} \qquad (4.1)$$

The latter is often termed conditional likelihood by ignoring the marginal density. Additionally, by the Markovian property, it suffices to study transition density of lag one. In fact, let $p_k(x|x')$ be the $k$-step transition density. Then $p_k(x|x') = \int p_1(x''|x')p_{k-1}(x|x'')dx''$. Therefore, $p_k$ can be obtained recursively from the one-step transition density $p_1$.

## *4.2. Random samples and high-frequency financial data*

In contrast to low-frequency financial data that are sampled regularly on a daily, weekly or monthly basis, high-frequency financial data are usually sampled at irregular random times. Let $0 = \tau_0 < \tau_1 < \cdots < \tau_{N_T} \leq T$ be $N_T + 1$ discrete observations up to time $T$. Let $\Delta_i = \tau_i - \tau_{i-1}$ be the sampling intervals. In low-frequency setting, $\Delta_i = \Delta$ is assumed to be constant. In high-frequency framework, $\Delta_i, i = 1, 2, \ldots,$ are assumed to be random variables. In fact, most real financial market transactions arrive irregularly and randomly. Among many examples are credit card purchases and stock buy/sale transactions.

To study the transaction times, Engle and Russell [70] propose a class of autoregressive conditional duration model given by

$$\Delta_i = \psi_i \varepsilon_i, \quad \psi_i = \mathbb{E}(\Delta_i|\Delta_{i-1}, \Delta_{i-2}, \ldots, \Delta_1), \tag{4.2}$$

where $\varepsilon_i, i = 1, 2, \ldots$, are iid random variables independent of $\psi_i$. Depending on different specifications, (4.2) includes many examples, including, for instances, $m$-memory model $\psi_i = \gamma + \sum_{j=0}^{m} \alpha_j \Delta_{i-j}$ and ACD$(m,q)$ model $\psi_i = \gamma + \sum_{j=0}^{m} \alpha_j \Delta_{i-j} + \sum_{j=0}^{q} \beta_j \psi_{i-j}$. When $\varepsilon_i$ follows standard exponential distribution, the latter model is termed exponential ACD (EACD) model. See [70] for more details. Zhang *et al.* [169] further extend the linear ACD model to threshold ACD (TACD) model by allowing the coefficients in the ACD model to vary according to the behavior of a threshold variable. Zhang *et al.* [169] use generalized Gamma distribution for $\varepsilon_i$, resulting in GACD model, and find strong evidence that stock dynamics behavior differently during fast transaction periods and slow transaction periods.

In addition to the randomness introduced by the underlying process, the randomness from the random sampling also plays an important role in statistical inferences. In the presence of random sampling, one needs to consider likelihoods for bivariate observations $(X_i, \Delta_i)$. Aït-Sahalia and Mykland [9] argue that the loss from not using sampling intervals is even greater than the loss due to the discreteness of samples. Duffie and Glynn [63] study random samples from a general Markov process.

### 4.3. Error-in samples

In practice, we may not observe $X_t$ directly but a contaminated version $X_t^*$ of it. For example, assume that $X_i$ is the actual stock returns based on logarithm during the $i$-th time period, and $X_i^*$ is the observed returns with errors. This phenomenon is closely related to the market microstructure and becomes more pronounced under high-frequency setting. For example, for a continuous semimartingale, it is well-known that the realized volatility computed using discrete observations converges in probability to the quadratic variation of the semimartingale as the sampling frequency increases. This, however, contradicts with empirical observations that realized volatility using high-frequency data generally does not stabilize; see Brown [47]. One possible explanation for this phenomenon is that the underlying process is contaminated with market microstructure errors.

There are two popular market microstructure error models in the literature: additive errors and rounding errors. The additive errors model assumes that $X_i^* = X_i + \xi_i$ with error $\xi_i$. The errors $\{\xi_i\}$ are assumed to be iid and independent of $\{X_i\}$. See, for example, Aït-Sahalia *et al.*[10], Hansen and Lunde [96], Zhang [167], Zhang *et al.* [168], and Zhou [175] to name a few. For rounding errors model, $X_i^*$ is taken to be the nearest multiple of a smallest unit $\alpha$ (say, 1 cent in stock prices). That is, $X_i^* = \alpha[X_i/\alpha]$, where $[\cdot]$ denotes the integer rounding operation. See Delattre and Jacod [59] and Zeng [166]. Li and Mykland [126] study a more general error model via Markov kernel:

$$\mathbb{P}(X_{t_i}^* \leq x | \{X_t\}_{t \geq 0}) = \mathbb{P}(X_{t_i}^* \leq x | X_{t_i}) = Q(X_{t_i}, x). \qquad (4.3)$$

That is, given $\{X_t\}_{t \geq 0}$, the contaminated version $X_i^*$ only depends on $X_{t_i}$.

Observations with contaminated errors make statistical inferences more difficult in a few aspects. Let us consider the additive errors $Y_i^* = X_i + \xi_i$. First, the contaminated process $\{X_i^*\}_{i \geq 0}$ may not form a Markov chain even if the original series do. Instead, it becomes a hidden Markov chain with the hidden chain $\{X_i\}_{i \geq 0}$; see Section 4.4. For likelihood based methods, we need to integrate out the unobservable process $\{X_i\}_{i \geq 0}$. Second, working with contaminated observations is essentially a deconvolution problem: extracting information about $X_i$ based on $X_i^*$. The latter problem is usually quite difficult. For example, nonparametric kernel density estimators for density function of $X_i$ have very slow rate of convergence; see Stefanski and Carroll [151], Liu and Taylor [130], and Fan [74]. Third, volatility computed from $\{X_i^*\}_{i \geq 0}$ has two components: volatility from the true process $\{X_i\}_{i \geq 0}$ and the errors $\{\xi_i\}_{i \geq 0}$. The latter term represents the bias and needs to be taken care of; see Aït-Sahalia *et al.* [10], Zhang [167], and Zhang *et al.* [168].

### *4.4. Hidden Markov models*

The Markov chain assumption works well for deterministic volatility models, that is, the volatility $\sigma$ is a deterministic function of the state variable $X_t$. Examples include the continuous-time model (2.7) and the nonlinear autoregressive model (3.1). In many applications, however, the Markov chain assumption is too restrictive. For example, in (2.12), let $X_i = \log(S_{i\Delta}) - \log(S_{(i-1)\Delta})$ be the aggregated log returns during time period $[(i-1)\Delta, i\Delta]$. Because the volatility $\sigma_t^2$ itself is an unobserved stochastic process with serial dependence, $\{X_i\}_{i \geq 0}$ does not form a Markov chain. Similarly, $\{X_i\}_{i \geq 0}$ from the discrete-time stochastic volatility model (3.5) is not Markovian. For stochastic volatility models, hidden Markov models (HMM) offer a good alternative; see Genon *et al.* [90] and Zhao [170]. Following Bickel and Ritov [37], Leroux [125] and Zhao [170], we give a definition of HMM.

**Definition 1.** *A stochastic process $\{X_i\}_{i \in \mathbb{Z}}$ with state space $(\mathbb{R}, \mathcal{B}(\mathbb{R}))$ is a hidden Markov model with respect to the hidden chain $\{Y_i\}_{i \in \mathbb{Z}}$ with state space $(\mathcal{Y}, \mathcal{B}(\mathcal{Y}))$ if*

(i) *$\{Y_i\}_{i \in \mathbb{Z}}$ is a strictly stationary Markov chain.*
(ii) *For all $i$, given $\{Y_j\}_{j \leq i}$, $\{X_j\}_{j \leq i}$ are conditionally independent, and the conditional distribution of $X_i$ depends only on $Y_i$.*
(iii) *The conditional distribution of $X_i$ given $Y_i = y$ does not depend on $i$.*

If $\{X_i\}_{i \in \mathbb{Z}}$ itself is a stationary Markov chain, then it is also a HMM with respect to the observable Markov chain $\{Y_i = X_{i-1}\}_{i \in \mathbb{Z}}$. Therefore, HMM includes

Markov chain as a special case. Zhao [170] has shown that, many continuous-time and discrete-time models used in financial econometrics are special examples of HMM. For example,

- In (2.7), $\{X_i\}$ is a HMM with respect to the observable chain $\{Y_i = X_{i-1}\}$.
- In (2.12), the aggregated log returns $\{X_i = \log(S_{i\Delta}) - \log(S_{(i-1)\Delta})\}$ is a HMM with respect to the unobservable chain $\{Y_i^{(1)} = (\sigma_{t+(i-1)\Delta}^2)_{t \in [0,\Delta]}\}$ or $\{Y_i^{(2)} = (\sigma_{(i-1)\Delta}^2, \int_{(i-1)\Delta}^{i\Delta} \sigma_t^2 dt)\}$;
- In (3.5), $\{X_i\}$ is a HMM with respect to the unobservable Markov chain $\{Y_i = \sigma_i\}$.
- In (3.3) and (3.4), $\{X_i\}$ is a HMM with respect to the observable chain $\{Y_i\}$.

HMM can also be used to describe observations with contaminated errors in Section 4.3. For example, consider the additive error model $X_i^* = X_i + \xi_i$, where the errors $\{\xi_i\}_{i \geq 0}$ are assumed to be iid and independent of $\{X_i\}_{i \geq 0}$. Clearly, if $\{X_i\}_{i \geq 0}$ is a Markov chain, then $\{X_i^*\}_{i \geq 0}$ is a HMM with respect to the unobservable chain $\{X_i\}_{i \geq 0}$. Examples satisfying this condition include models (2.7), (3.1), and the hyperbolic Lévy motion model in [67]. On the other hand, the HMM structure may still hold even if $\{X_i\}_{i \geq 0}$ does not form a Markov chain. For example, consider the stochastic volatility model (3.5), then $\{X_i\}_{i \geq 0}$ is not a Markov chain, but $\{X_i^*\}_{i \geq 0}$ is still a HMM with respect to the unobservable chain $\{Y_i = \sigma_i\}_{i \geq 0}$ provided that $\eta_i$ and $\varepsilon_i$ in (3.5) and the errors $\xi_i$ are independent. A similar statement holds true for the continuous-time stochastic volatility model in Section 2.3. Genon *et al.* [90] and Zhao [170] also show that certain dependence structure (for example, mixing properties) of the hidden chain $\{Y_i\}$ carries over to $\{(X_i, Y_i)\}$, and hence many tools for Markov chain are also applicable to HMM.

## 5. Model estimation: parametric methods

In this section we review some popular parametric estimation methods in financial econometrics. When we have a sufficient amount of prior information about the underlying model, for example, the model is from a parametric family $\{\mathcal{M}_\theta, \theta \in \Theta\}$, where $\mathcal{M}_\theta$ is a known parametric form with unknown parameter $\theta$, then the main focus becomes the estimation of the parameter $\theta$. Parametric methods address estimation problems in such contexts.

### 5.1. Likelihood based method

Given observations $\{X_i\}_{0 \leq i \leq n}$, if we know the parametric form of the model that generates $\{X_i\}_{0 \leq i \leq n}$, then maximum likelihood is the natural method. Suppose that $\{X_i\}_{i \geq 0}$ form a stationary Markov chain with invariant density $\pi(x; \theta)$ and transition density $p(x|x';\theta)$. Then the log likelihood function is

$$\begin{aligned}\ell(X_0,\ldots,X_n;\theta) &= \sum_{i=1}^n \log[p(X_i|X_{i-1};\theta)] + \log[\pi(X_0;\theta)] \\ &\approx \sum_{i=1}^n \log[p(X_i|X_{i-1};\theta)].\end{aligned} \quad (5.1)$$

The maximum likelihood estimate (MLE) is $\hat{\theta} = \mathrm{argmax}_\theta\, \ell(X_0,\ldots,X_n;\theta)$. Consider, for example, model (2.1). Let $X_i = \log(S_{i\Delta}) - \log(S_{(i-1)\Delta})$ be the aggregated returns during $[(i-1)\Delta, i\Delta]$. Then $\{X_i\}_{1\le i\le n}$ are iid normal random variables with mean $(\mu - \sigma^2/2)\Delta$ and variance $\sigma^2\Delta$. Therefore, we have explicit form in (5.1). For Vasicek model (2.3), $\{X_i = r_{i\Delta}\}_{i\ge 0}$ have Gaussian transition density with mean $(1-\rho)\alpha + \rho X_{i-1}$ and variance $\sigma^2(1-\rho^2)/(2\beta)$, where $\rho = \exp(-\beta\Delta)$. For the CIR model (2.4), the transition density is a noncentral chi-square distribution with parameters fully determined by $\alpha, \beta$ and $\sigma$. It is also easy to write down the transition density for many parametric time series models. For example, in (3.1), let $\varepsilon_i$ be iid standard normals. Denote by $\phi(x)$ the standard normal density. Then $p(X_i|X_{i-1};\theta) = \phi\{[X_i - \mu_\theta(X_{i-1})]/\sigma_\theta(X_{i-1})\}/\sigma_\theta(X_{i-1})$. For many continuous-time models, however, one practical issue arises. Except for models (2.1), (2.3) and (2.4), transition densities for many other parametric models do not admit closed forms. One way out is to use the following Euler approximation scheme for (2.7):

$$X_{t+\Delta} = X_t + \mu(X_t)\Delta + \sigma(X_t)\Delta^{1/2}\varepsilon_t, \quad t = 0, \Delta, 2\Delta, \ldots, \quad (5.2)$$

where $\{\varepsilon_{i\Delta}\}_{i\ge 0}$ are iid standard normals. In fact, most continuous-time models used in finance are estimated based on the approximation (5.2). The approximation works well in high-frequency setting. An alternative approach is the approximation method in Aït-Sahalia [3; 4] where the likelihood is approximated by a sequence of likelihoods based on Hermite polynomials. See the survey paper by Aït-Sahalia [5] on likelihood methods for (2.8) and its multivariate version.

For hidden Markov models in Section 4.4, since the hidden chain $\{Y_i\}_{0\le i\le n}$ is not observable, we need to integrate out $\{Y_i\}_{0\le i\le n}$ in order to obtain the likelihood function for the observations $\{X_i\}_{0\le i\le n}$. To be precise, denote by $f(x|y;\theta)$ the conditional density of $X_i$ at $x$ given $Y_i = y$, by $Q(\cdot|y;\theta)$ the transition probability measure of $Y_i$ given $Y_{i-1} = y$, and by $Q(\cdot;\theta)$ the invariant probability measure of $Y_0$. Then the likelihood function for observations $\{X_i\}_{0\le i\le n}$ is

$$L(X_0,\ldots,X_n;\theta) = \int \prod_{i=0}^n f(X_i|y_i;\theta)Q(dy_0;\theta)\prod_{i=1}^n Q(dy_i|y_{i-1};\theta). \quad (5.3)$$

Due to the high dimensional integral, direct computation and maximization of $L(X_0,\ldots,X_n;\theta)$ is computationally infeasible. This makes estimation of stochastic volatility models quite difficult; see Section 5.4.

### *5.2. Generalized method of moments*

The generalized method of moments [GMM, Hansen [95]] is a popular parameter estimation method in finance. Assume that we have a stationary process $\{X_t\}_{t\geq 0}$ whose data-generating mechanism involves parameter $\theta$. The essential idea of GMM works as follows:

(a) Derive a set of theoretical moments conditions. That is, for properly chosen function $g_\theta$, find constant $C_{g,\theta}$ such that

$$\mathbb{E}[g_\theta(X_0)] = C_{g,\theta}, \quad \text{or} \quad \mathbb{E}[\tilde{g}_\theta(X_0)] = 0 \text{ with } \tilde{g}_\theta = g_\theta - C_{g,\theta}. \tag{5.4}$$

(b) Minimize certain measure of the discrepancies between the empirical and the theoretical moments. Namely, for a chosen criterion norm $\delta$,

$$\hat{\theta} = \operatorname*{argmin}_{\theta} \delta\Big[\frac{1}{n}\sum_{i=1}^n \tilde{g}_\theta(X_i)\Big] = \operatorname*{argmin}_{\theta} \delta\Big[\sum_{i=1}^n \tilde{g}_\theta(X_i)\Big], \tag{5.5}$$

For example, if $\delta(u) = u^2$, then we have least-squares type estimate. In (5.5), one often uses weighted discrepancies for a set of functions $\tilde{g}_\theta$. For large sample properties of GMM, see Hansen [95].

The key step in GMM is step (a). We now introduce the idea in Hansen and Scheinkman [97] to derive moments conditions for a stationary Markov process $\{X_t\}_{t\geq 0}$. Let $\{\mathcal{J}_t\}_{t\geq 0}$ be a family of operators defined by $\mathcal{J}_t g(x) = \mathbb{E}[g(X_t)|X_0 = x]$. Notice that $\mathcal{J}_0 g(x) = g(x)$. The operators $\{\mathcal{J}_t\}_{t\geq 0}$ uniquely determine the transition density of $X_t$ given $X_0$ by taking $g(x) = \exp(iux), u \in \mathbb{R}$. Introduce the infinitesimal generator [see Karlin and Taylor [117] and Hansen and Scheinkman [97]] of $X_t$, $\mathcal{L}$, given by

$$\mathcal{L}g(x) = \frac{\partial \mathcal{J}_t g(x)}{\partial t}\Big|_{t=0} = \lim_{t\downarrow 0}\frac{\mathcal{J}_t g(x) - g(x)}{t}. \tag{5.6}$$

By stationarity, $\mathbb{E}[\mathcal{J}_t g(X_0)] = \mathbb{E}[g(X_0)]$. Therefore, assuming that we can exchange the order of expectation and differentiation, (6.6) implies that

$$\mathbb{E}[\mathcal{L}g(X_0)] = 0. \tag{5.7}$$

The expression (5.7) holds for all functions $g$ satisfying some regularity conditions. Thus, we can, in principle, produce infinitely many moments conditions.

By Itô's formula, the infinitesimal generator of $\{X_t\}_{t\geq 0}$ from the stochastic differentiation equation (2.7) is given by

$$\mathcal{L}g(x) = \frac{\partial g(x)}{\partial x}\mu(x) + \frac{\partial^2 g(x)}{\partial x^2}\frac{\sigma^2(x)}{2}. \tag{5.8}$$

So, for a given parametrization $(\mu, \sigma) = (\mu_\theta, \sigma_\theta)$, we can use (5.8) and GMM to estimate $\theta$; see Hansen and Scheinkman [97]. Duffie and Glynn [63] apply GMM to estimate parameters based on random samples from a Markov process.

Condition (5.7) only uses information from marginal stationarity. A more efficient approach would incorporate transition or conditional information into moments conditions. Since most transition densities except a few rare cases do not have explicit form, any approach relying on transition information needs some approximation technique. For example, for the CKLS model (2.5), we can apply the Euler approximation scheme to obtain approximated moments conditions:

$$\mathbb{E}(r_t) = \alpha, \quad \mathbb{E}(\epsilon_{t+\Delta}|r_t) \approx 0, \quad \mathbb{E}[\epsilon_{t+\Delta}^2|r_t] \approx \sigma^2 r_t^{2\gamma}\Delta,$$

where $\epsilon_{t+\Delta} = r_{t+\Delta} - r_t - \beta(\alpha - r_t)\Delta$.

In (5.4), the assumption that $C_{g,\theta}$ needs to be of a known form limits significantly the applicability of the GMM estimator. To overcome this difficulty, Duffie and Singleton [61] introduce a simulated moments estimation method which estimates the parameters of interest by matching the sample moments of the actual and simulated process.

### 5.3. Other parameter estimation methods for diffusion models

We briefly mention other parameter estimation methods for model (2.8). Density based parameter estimation is to minimize discrepancy between the nonparametric density estimate and the theoretical parametric density or its parametric estimate. Consider, for example, model (2.7). Under parametric setting $(\mu, \sigma) = (\mu_\theta, \sigma_\theta)$, the theoretical stationary density of $\{X_t\}$ is given by $f_\theta = f_{\mu_\theta, \sigma_\theta}$ in (6.1). Let $\hat{f}$ be the nonparametric kernel density estimate in (6.2). Then $\theta$ can be estimated by

$$\hat{\theta} = \operatorname*{argmin}_{\theta} \sum_{i=1}^n [\hat{f}(X_i) - f_\theta(X_i)]^2. \tag{5.9}$$

Aït-Sahalia [2] establishes $\sqrt{n}$-consistency for $\hat{\theta}$.

Other contributions include martingale estimation function method in Bibby and Sørensen [35] and Kessler and Sørensen [118] among others.

### 5.4. Parameter estimations in stochastic volatility models

As argued in Section 5.1, it is computationally infeasible to estimate parameters in stochastic volatility models using direct maximum likelihood methods. Here we briefly review some alternatives. One popular approach is various moment based methods in, for example, Andersen and Lund [16], Andersen and Sørensen [17], Gallant and Tauchen [86], Melino and Turnbull [134], Taylor [152], and Wiggins [158]. The basic idea is to express the parameters of interest in terms of population (conditional) moments and replace the latter by sample (conditional) moments, or employ the generalized methods of moments (GMM) in Section 5.2. Andersen and Sørensen [17] study the finite sample performance of GMM estimation of the stochastic volatility model (3.6). Andersen *et al.* [15] examine small-sample properties of the efficient method of moments proposed by Bansal *et al.* [26] and Gallant and Tauchen [86]. Other methods include quasi-maximum likelihood in Ruiz [144] and Harvey *et al.* [100], Bayesian Markov-chain Monte Carlo method in Chib *et al.* [50], Jacquier *et al.* [108; 109] and Kim *et al.* [119], and the method in Genon *et al.* [89]. See the survey paper by Broto and Ruiz [46].

## 6. Model estimation: nonparametric methods

### 6.1. Nonparametric density estimates

One important goal of financial econometrics is to study the distribution of returns from financial markets. Such distribution can provide rich information about the underlying process driving the financial markets. For example, Mandelbrot [132] finds that cotton price changes have heavy tails relative to normal distributions. This motivated him to use stable distribution as a possible alternative over the traditional normal distribution. For the past three decades, the leptokurtic property and volatility smile observed in financial data have been the driven force for searching for more appropriate models than the Black-Scholes model to account for empirical characteristics of financial data.

To appreciate the idea more, consider model (2.7). Let $f$ be the marginal density function of the stationary solution $X_t$ on $D = (D_l, D_u)$ with $-\infty \leq D_l < D_u \leq +\infty$. Under some regularity conditions, $f$ is given by

$$f(x) = \frac{c(x_0)}{\sigma^2(x)} \exp\Big\{ \int_{x_0}^{x} \frac{2\mu(y)}{\sigma^2(y)} dy \Big\}, \qquad (6.1)$$

where the choice of the lower bound point $x_0 \in D$ is irrelevant, and $c(x_0)$ is a normalizing constant to ensure that $f$ is a probability density on $D$; see Aït-Sahalia [2]. Therefore, the marginal density $f$ has an intrinsic connection to the drift $\mu$ and the diffusion $\sigma$, which can be used to do model validation or model parameter estimation. For example, Aït-Sahalia [2] and Zhao [170] study model validation problem $H_0 : (\mu, \sigma) = (\mu_\theta, \sigma_\theta)$ for model (2.7) by comparing nonparametric density estimate and parametric density estimate under $H_0$; see Section 7.6. Aït-Sahalia [1] constructs nonparametric estimate of the diffusion function $\sigma$ through nonparametric density estimate.

Let $\{X_t\}_{t \in \mathcal{T}}$ be a generic stationary process. For example, it could be the continuous-time process (2.7), discrete-time process including aggregated returns in (2.12), nonlinear time series in Section 3.1, among others. Given discrete observations $\{X_i\}_{1 \leq i \leq n}$, the classical nonparametric kernel density estimate for the density $f$ of $X_1$ is given by

$$\hat{f}(x) = \frac{1}{nb_n} \sum_{i=1}^{n} K_{b_n}(X_i - x), \quad \text{where} \quad K_{b_n}(u) = K(u/b_n). \qquad (6.2)$$

See Silverman [147]. Wu and Mielniczuk [162] study the asymptotic behavior of $\hat{f}$ for linear process. Assuming some mixing conditions, it is possible to establish asymptotic normality for $\hat{f}$ with optimal rate $O(n^{-2/5})$.

Recently, Schick and Wefelmeyer [145] and Kim and Wu [120] propose convolution-type efficient density estimates that achieve the parametric rate $O(n^{-1/2})$. To appreciate the idea, consider the model

$$X_i = \mu_\theta(X_{i-1}) + \varepsilon_i, \tag{6.3}$$

where $(\varepsilon_i)_{i\in\mathbb{Z}}$ are iid and $\mathbb{E}(\varepsilon_0) = 0$, and $\mu_\theta$ is a known parametric form with unknown parameter $\theta$. The popular AR, TAR, and EAR models are of form (6.3). The ARCH model $X_i = \sqrt{a^2 + b^2 X_{i-1}^2}\varepsilon_i$ is also of form (6.3) after transformation: $\log(X_i^2) = \log(a^2 + b^2 X_{i-1}^2) + \log(\varepsilon_i^2)$. Denote by $f_{\mu_\theta(X)}$ and $f_\varepsilon$ the density functions of $\mu_\theta(X_0)$ and $\varepsilon_0$, respectively. By convolution,

$$f(x) = \int f_{\mu_\theta(X)}(y) f_\varepsilon(x-y) dy. \tag{6.4}$$

Then the convolution-type estimate procedure works as follows:

(a) Obtain a $\sqrt{n}$-consistent estimate $\hat{\theta}$ of $\theta$ by least-squares method or M-estimation method.
(b) Compute $\mu_{\hat{\theta}}(X_{i-1})$ and $\hat{\varepsilon}_i = X_i - \mu_{\hat{\theta}}(X_{i-1}), 1 \leq i \leq n$.
(c) Obtain nonparametric kernel density estimates $\hat{f}_{\mu_\theta(X)}$ and $\hat{f}_\varepsilon$ of $f_{\mu_\theta(X)}$ and $f_\varepsilon$, respectively, by the estimated values $\mu_{\hat{\theta}}(X_{i-1})$ and $\hat{\varepsilon}_i$ via (6.2).
(d) In (6.4), replace $f_{\mu_\theta(X)}$ and $f_\varepsilon$ by their estimates, and obtain $\hat{f}$.

Kim and Wu [120] establish a $\sqrt{n}$ central limit theorem for the resulting convolution-type estimate $\hat{f}$. Schick and Wefelmeyer [145] obtain a similar result for linear process. Zhao [171] studies efficient density estimation for conditional heteroscedastic models.

Since the distributional property of a stationary Markov process can be characterized by the marginal and transition density functions, let us now consider the transition density of a Markov process $\{X_t\}_{t\in\mathcal{T}}$ based on discrete observations $\{X_i\}_{1\leq i\leq n}$. By the Markovian property, it suffices to consider the transition density at time lag one, that is, the conditional density function $\pi(x|x')$ of $X_i$ at $x$ given that $X_{i-1} = x'$. Denote by $\pi(x, x')$ and $f(x)$ the joint density of $(X_0, X_1)$ and the marginal density of $X_0$, respectively. Since $\pi(x|x') = \pi(x, x')/f(x)$, a nonparametric estimate of $\pi(x|x')$ can be constructed by plugging in the nonparametric estimates of the latter two densities.

### *6.2. Nonparametric function estimation*

In contrast to parametric methods, nonparametric methods on function estimation do not assume any parametric form of the function other than certain smoothness assumption. Suppose, for instance, that we want to estimate the mean regression function $\mathbb{E}(X_i|Y_i = y)$. Let $f(y)$ and $f(x,y)$ be the densities of $Y_1$ and $(X_1, Y_1)$, respectively. Then we have

$$\mathbb{E}(X_i|Y_i = y) = \frac{1}{f(y)} \int x f(x,y) dx \approx \frac{\sum_{i=1}^n X_i K_{b_n}(x - Y_i)}{\sum_{i=1}^n K_{b_n}(x - Y_i)}. \quad (6.5)$$

Expression (6.5) can be used to construct nonparametric estimates for drift and volatility functions in financial models. Consider model (3.3), assume that $\mathbb{E}(\varepsilon_0) = 0$ and $\mathbb{E}(\varepsilon_0^2) = 1$. Then $\mu(y) = \mathbb{E}(X_i|X_i = y)$ and $\sigma^2(y) = \mathbb{E}\{[X_i - \mu(Y_i)]^2|Y_i = y\}$. The idea is as follows:

(a) Apply (6.5) to get a nonparametric estimate $\hat{\mu}$ of $\mu$.
(b) Compute residuals $\hat{\varepsilon}_i = X_i - \hat{\mu}(Y_i)$.
(c) Apply (6.5) to $(\hat{\varepsilon}_i^2, Y_i)$ to nonparametrically estimate $\sigma^2$.

See Fan and Yao [76] and Zhao and Wu [174] for related works. The latter papers also show that the volatility function can be estimated as well as if we know the drift function. The intuition is that the bias term resulting from estimating the drift is of order $O(b_n^2)$ and is squared to $O(b_n^4)$ when estimating the volatility.

To apply the above idea to the estimation of the continuous-times model (2.7), we follow Stanton [149] and introduce the infinitesimal generator of $X_t$, $\mathcal{L}$, given by

$$\begin{aligned}\mathcal{L}g(x,t) &= \lim_{\Delta \downarrow 0} \frac{\mathbb{E}[g(X_{t+\Delta}, t+\Delta)|X_t = x] - g(x,t)}{\Delta} \\ &= \frac{\partial g(x,t)}{\partial t} + \frac{\partial g(x,t)}{\partial x}\mu(x) + \frac{\partial^2 g(x,t)}{\partial x^2}\frac{\sigma^2(x)}{2}\end{aligned} \quad (6.6)$$

in view of Itô's formula. As a special case, if the function $f$ does not depend on $t$, then (6.6) reduces to (5.8). Apply a Taylor's expansion to (6.6),

$$\mathbb{E}[g(X_{t+\Delta}, t+\Delta)|X_t] = g(X_t, t) + \mathcal{L}g(X_t, t)\Delta + \frac{1}{2}\mathcal{L}^2 g(X_t, t)\Delta^2 + \cdots. \quad (6.7)$$

Thus, a first order approximation of $\mathcal{L}g(X_t, t)$ is

$$\mathcal{L}g(X_t, t) = \frac{1}{\Delta}\mathbb{E}\{[g(X_{t+\Delta}, t+\Delta) - g(X_t, t)]|X_t\} + O(\Delta). \quad (6.8)$$

Taking $g(x,t) = x$, then $\mathcal{L}g(X_t, t) = \mu(X_t)$ and

$$\mu(x) = \frac{1}{\Delta}\mathbb{E}[(X_{t+\Delta} - X_t)|X_t = x] + O(\Delta). \quad (6.9)$$

Similarly, taking $g(x,t) = (x - X_t)^2$, we have

$$\sigma^2(x) = \frac{1}{\Delta}\mathbb{E}[(X_{t+\Delta} - X_t)^2|X_t = x] + O(\Delta). \quad (6.10)$$

Therefore, (6.9) and (6.10) can be used to construct nonparametric estimates of $\mu$ and $\sigma$ in conjunction with (6.5).

For the simplified model $dX_t = \sigma(X_t)dW_t$, Arfi [18] uses (6.10) to estimate $\sigma^2$ via kernel smoothing. Other contributions on nonparametric estimation of (2.7) include Bandi [23], Bandi and Phillips [24], Florens-Zmirou [82], Foster and Nelson [83], and Jiang and Knight [112]. Higher-order approximations are considered in Stanton [149] to reduce biases. Fan and Zhang [78], however, argue that the bias reduction is achieved at the cost of exponential increase of the variance. Therefore, they suggest that one should avoid using too higher-order approximations in practice.

For the continuous-times model (2.7), an alternative approach is based on the Euler approximation scheme (5.2). Then we can estimate $\mu$ and $\sigma$ through the expressions:

$$\mathbb{E}[\Delta^{-1}(X_{t+\Delta} - X_t)|X_t = x] = \mu(x);$$
$$\mathbb{E}[\Delta^{-1}(X_{t+\Delta} - X_t - \mu(X_t)\Delta)^2|X_t = x] = \sigma^2(x).$$

For (5.2) to approximate (2.7) with a reasonable accuracy, $\Delta$ needs to be very small. Therefore, (5.2) is often useful in dealing with high-frequency data over a long time span: $\Delta = \Delta_n \to 0$ and $n\Delta_n \to \infty$.

The estimate (6.5) is basically a local constant fit based on weighted least-squares. There are two natural variants. The first one is the local linear method. Let $\mu(y) = \mathbb{E}(X_i|Y_i = y)$. The local linear estimate of $(\mu(y), \mu'(y))$ is

$$(\hat{\mu}(y), \hat{\mu}'(y)) = \operatorname*{argmin}_{(a,b)} \sum_{i=1}^n [X_i - a - b(Y_i - y)]^2 K_{b_n}(y - Y_i). \qquad (6.11)$$

Local linear estimates can reduce boundary effect. For model (3.3), Fan and Yao [76] use (6.11) to estimate $\mu$ first, and then apply (6.11) to the squared residuals $[X_i - \hat{\mu}(Y_i)]^2$ to estimate $\sigma^2$.

Another variant of (6.5) is the least-absolute-deviation (LAD) estimate. In model (3.3), assume that median$(\varepsilon_0) = 0$ and median$|\varepsilon_0| = 1$. Then $\mu(y) =$ median$(X_i|Y_i = y)$ and $\sigma(y) =$ median$[|X_i - \mu(Y_i)||Y_i = y]$. Thus, the LAD estimates of $\mu$ and $\sigma$ are

$$\hat{\mu}(y) = \operatorname*{argmin}_{\mu} \sum_{i=1}^n \left|X_i - \mu\right| K_{b_n}(y - Y_i), \qquad (6.12)$$

$$\hat{\sigma}(y) = \operatorname*{argmin}_{\sigma} \sum_{i=1}^n \left||X_i - \hat{\mu}(Y_i)| - \sigma\right| K_{b_n}(y - Y_i). \qquad (6.13)$$

Basically, LAD estimate is a local median type estimate and hence it is robust against outliers. Under a very general dependence structure, Zhao and Wu [172] study the asymptotic properties of the LAD estimates for (3.3). The results obtained are applicable to a variety of time series models, including linear processes and nonlinear models (3.1) and (3.3).

### 6.3. Semi-parametric estimation via nonparametric density

Semi-parametric models are between parametric and nonparametric ones by imposing parametric form on part of the model while keeping other parts nonparametric. They are particularly useful when we have prior knowledge about part of the data generating process while staying flexible on the remaining parts. For example, consider model (2.7). We have two semi-parametrizations: (i) $\mu = \mu_\theta$ for a known parametric form $\mu_\theta$ with unknown parameter $\theta$; and (ii) $\sigma = \sigma_\theta$ for a known parametric form $\sigma_\theta$ with unknown parameter $\theta$. The two frameworks have different ranges of applicability. Many empirical studies suggest fitting a simple form for $\mu$. For example, [1; 48; 53; 156] use linear form for $\mu$, and [2] uses nonlinear drift form (2.6) to fit interest rates. For high-frequency data (say, daily, hourly, or 5-minute), it is even reasonable to assume $\mu$ to be constant or zero since we are more interested in the volatility instead. Volatility is very important in options pricing. Options written on volatile assets are more expensive. In such circumstances, it is desirable to treat the volatility nonparametrically to avoid mis-specification. On the other hand, in some cases it may be reasonable to assume a parametric form for the volatility function while keeping the drift nonparametric; see Kristensen [123] and Banon [25]. For an extensive exposition of semi-parametric methods, see Gao [87].

Here, we briefly review the estimation methods in [1] and [25] for the two semi-parametrizations of model (2.7). Let $p(\Delta, x|x')$ be the transition density of $X_{t+\Delta}$ at $x$ given $X_t = x'$, and $\pi(x)$ the stationary density of $X_t$. Then the Kolmogorov forward equation associated with (2.7) is

$$\frac{\partial p(\Delta, x|x')}{\partial \Delta} = -\frac{\partial [\mu(x) p(\Delta, x|x')]}{\partial x} + \frac{\partial^2 [\sigma^2(x) p(\Delta, x|x')]}{2 \partial x^2}. \tag{6.14}$$

By stationarity $\int p(\Delta, x|x')\pi(x')dx' = \pi(x)$. Multiply (6.14) by $\pi(x')$ and take integral with respect to $x'$,

$$\frac{d^2[\sigma^2(x)\pi(x)]}{dx^2} = 2\frac{d[\mu(x)\pi(x)]}{dx}. \tag{6.15}$$

In Aït-Sahalia [1], he assumes $\mu = \mu_\theta$. Integrate (6.15) twice to obtain

$$\sigma^2(x) = \frac{2}{\pi(x)} \int_0^x \mu_\theta(u) \pi(u) du. \tag{6.16}$$

Let $\hat{\theta}$ be a consistent estimate of $\theta$, and $\hat{\pi}(\cdot)$ a nonparametric kernel density estimate constructed as in (6.2). Then we can plug $\hat{\theta}$ and $\hat{\pi}$ into (6.16) to obtain a nonparametric estimate of $\sigma^2(x)$. For $\mu_\theta(x) = \beta(\alpha - x), \theta = (\alpha, \beta)$, $\theta$ can be estimated through the regression equation $\mathbb{E}(X_{t+\Delta}|X_t) = \alpha + e^{-\beta\Delta}(X_t - \alpha)$. This approach has an apparent advantage: it always works regardless of the size of $\Delta$ while Stanton's [149] method requires high-frequency data $\Delta \to 0$. See [1] for more details. Banon [25] integrates (6.15) once to obtain

$$\mu(x) = \frac{1}{2\pi(x)} \frac{d[\sigma^2(x)\pi(x)]}{dx}. \tag{6.17}$$

Banon [25] considers a constant but unknown $\sigma$. The drift $\mu$ is nonparametrically estimated using (6.17) with estimated $\sigma$ and density $\pi(\cdot)$.

### 6.4. Nonparametric integrated volatility estimation

Most finance theory lies within semimartingale framework. A stochastic process $\{X_t\}_{t\geq 0}$ (assuming $X_0 = 0$) is said to be a continuous semimartingale with respect to a filtration process $\{\mathcal{F}_t\}_{t\geq 0}$ if

$$X_t = M_t + A_t, \quad t \geq 0, \tag{6.18}$$

where $\{A_t, \mathcal{F}_t\}_{t\geq 0}$ is an adapted process with bounded variation paths on any finite subinterval of $[0, \infty)$, and $\{M_t, \mathcal{F}_t\}_{t\geq 0}$ is a continuous local martingale; see p. 149 in Karatzas and Shreve [116]. Denote by $\langle X \rangle_t$ the quadratic variation process of $\{X_t\}_{t\geq 0}$—the unique adapted and increasing process such that $\langle X \rangle_0 = 0$ and $\{X_t^2 - \langle X \rangle_t\}$ is a martingale (cf. Doob-Meyer decomposition). For continuous semimartingale, $\langle X \rangle_t = X_t^2 - 2\int_0^t X_s dX_s$. The fundamental result states that

$$\lim_{\max_i |t_{i+1}-t_i| \to 0} \sum_{i=0}^{n-1} (X_{t_{i+1}} - X_{t_i})^2 = \langle X \rangle_t, \quad \text{in probability}, \tag{6.19}$$

for all partitions $0 = t_0 < t_1 < \cdots < t_{n-1} < t_n = t$. The left hand side is often called realized volatility.

The latter result has important implications in finance theory. Consider the general stochastic volatility model

$$d\log(S_t) = \mu_t dt + \sigma_t dW_t, \tag{6.20}$$

where the drift $\{\mu_t\}$ and the volatility $\{\sigma_t\}$ are two adapted stochastic processes. In this special case, it can be shown using Itô's Lemma that $\langle \log(S) \rangle_t = \int_0^t \sigma_s^2 ds$ is the integrated volatility (contrast to the spot volatility $\sigma_t$). Therefore, without assuming any structure on $\{\mu_t\}$ and $\{\sigma_t\}$ other than some regularity conditions, we can estimate the integrated volatility nonparametrically using the realized volatility

$$\int_0^t \sigma_s^2 ds \approx \sum_{0=t_0<t_1<\cdots<t_{n-1}<t_n=t} [\log(S_{t_{i+1}}) - \log(S_{t_i})]^2. \tag{6.21}$$

In practice, one can compute daily, weekly, or monthly realized volatility based on high-frequency (say, 5-minute) data. Research along this line has been initiated by Andersen and Bollserslev [12], and Barndorff-Nielsen and Shephard [28]. Barndorff-Nielsen and Shephard [29] further obtain the asymptotic normality of the realized volatility. See the survey paper by Barndorff-Nielsen and Shephard [30]. Recent contributions include Andersen *et al.* [13], Mykland and Zhang [137], Zhang [167], and Zhang *et al.* [168]. The latter two papers deal with integrated volatility estimation for noisy high-frequency data. Zhao and Wu [174] study integrated volatility estimation for Lévy processes.

## 7. Model validation

As we have discussed in Section 2 and Section 3, there has never been a lack of parametric models. Parametric models can provide parsimonious interpretation of the data generating mechanism underlying the process, yet this is true only when the parametric models are correctly specified. For any parametric model, there is always a mis-specification risk that could lead to wrong conclusions. In fact, a correct specification for the price of the underlying asset is particularly important for the pricing of derivatives written on that asset. Therefore, one has to validate the adequacy of the parametric model before applying it to real data. Suppose that $\mathcal{Q}$ is the unknown characteristic of interest behind the underlying data generating mechanism. For example, in (2.7) and (3.1), we may take $\mathcal{Q} = (\mu, \sigma)$. For a given specification $\mathcal{Q}_\theta$ with possible unknown parameter $\theta$, we want to test the null hypothesis $H_0 : \mathcal{Q} = \mathcal{Q}_\theta$. This problem is often termed specification testing, model validation, model checking, goodness-of-fit, among others. There is a huge amount of literature on model validations. In this section we review some popular model validation techniques. In the rest of this section we implicitly assume that the alternative hypothesis is that the relevant functions are fully nonparametric.

### 7.1. Pseudo-likelihood ratio test

We briefly introduce the pseudo-likelihood ratio test (PLRT) in Azzalini and Bowman [19]. The PLRT is to compare the pseudo-likelihoods under both the null and the alternative. To appreciate the idea, we consider the simple regression model $Y_i = \mu(X_i) + e_i$. Suppose that we are interested in testing the null hypothesis $H_0 : \mu(x) = a + bx$ for some $a, b \in \mathbb{R}$.

Under $H_0$, we have a simple linear regression problem. Denote by $\mathbf{X}$ and $\mathbf{Y}$ the design matrix and the vector of responses, respectively. Under $H_0$, the fitted values are $\hat{\mathbf{Y}}_0 = \mathbf{H}_0 \mathbf{Y}$, where $\mathbf{H}_0 = \mathbf{X}(\mathbf{X}^T\mathbf{X})^{-1}\mathbf{X}^T$. The the residuals vector $\mathbf{e} = \mathbf{Y} - \mathbf{H}_0 \mathbf{Y}$, and the residual sum of squares $\mathrm{RSS}_0 = \mathbf{e}^T\mathbf{e} = \mathbf{Y}^T(\mathbf{I} - \mathbf{H}_0)\mathbf{Y}$. Under the alternative hypothesis of nonparametric setting, $\mu$ can be estimated by nonparametric regression methods (6.5) or (6.11) in Section 6.2. The fitted values $\hat{\mathbf{Y}}_1 = \mathbf{H}_1 \mathbf{Y}$, where $\mathbf{H}_1$ is a weighting matrix depending on $\mathbf{X}$, and the residual sum of squares $\mathrm{RSS}_1 = \mathbf{Y}^T(\mathbf{I} - \mathbf{H}_1)^T(\mathbf{I} - \mathbf{H}_1)\mathbf{Y}$. Then a naive PLRT is given by

$$T = \frac{\mathrm{RSS}_0 - \mathrm{RSS}_1}{\mathrm{RSS}_1}.$$

Azzalini and Bowman [19] point out that the null distribution of $T$ depends on the linear coefficient $b$, which makes $T$ unsuitable for hypothesis testing. To overcome this difficulty, they view the residuals vector $\mathbf{e} = \mathbf{Y} - \mathbf{H}_0\mathbf{Y} = (\mathbf{I} - \mathbf{H}_0)\mathbf{Y}$ as the new responses vector and apply the PLRT to $\mathbf{e}$ instead of $\mathbf{Y}$. They propose the new test statistic based on $\mathbf{e}$:

$$T^* = \frac{\mathbf{e}^T\mathbf{e} - \mathbf{e}^T(\mathbf{I} - \mathbf{H}_1)^T(\mathbf{I} - \mathbf{H}_1)\mathbf{e}}{\mathbf{e}^T\mathbf{e}}.$$

A large value of $T^*$ indicates rejection of $H_0$. Under a normality assumption on the distribution of $\mathbf{e}$, the null distribution of $T^*$ is related to quadratic form of normal variables; see the references in [19]. It is clear that one can extend the idea to more general model validations settings, although the test statistic for nonlinear parametric models may not possess an explicit form as in the linear regression case; see [20] and [72] on related works.

### 7.2. Model validation via nonparametric curve regression

In a stimulating paper, Härdle and Mammon [99] introduce a nonparametric curve regression based model validation procedure. Because nonparametric curve estimate (for example, kernel smoothing) is always a consistent estimate of the function of interest regardless of the underlying model, it is therefore natural to compare the parametric curve estimate under the null to the nonparametric curve estimate. Consider $Y_i = \mu(X_i) + e_i$. The popular nonparametric estimate of $\mu$ is

$$\hat{\mu}(x) = \frac{\sum_{i=1}^n Y_i K_{b_n}(x - X_i)}{\sum_{i=1}^n K_{b_n}(x - X_i)}. \tag{7.1}$$

Under $H_0: \mu = \mu_\theta$, we apply parametric methods to obtain a consistent estimate of $\theta$, denoted by $\hat{\theta}$. To mimic the structure of (7.1), we obtain the following parametric estimate

$$\hat{\mu}_{\hat{\theta}}(x) = \frac{\sum_{i=1}^n \mu_{\hat{\theta}}(X_i) K_{b_n}(x - X_i)}{\sum_{i=1}^n K_{b_n}(x - X_i)}. \tag{7.2}$$

Härdle and Mammon [99] use an $L_2$ distance between $\hat{\mu}$ and $\hat{\mu}_{\hat{\theta}}$ as the test statistic for $H_0$. See also Horowitz and Spokoiny [104].

An alternative approach is to compare the residuals sum squares under the null parametric model to that using nonparametric model. See Hong and White [103] along this line.

### 7.3. Residuals based test

A good statistical model would make the residuals behave like white noises. Therefore, a natural method of model checking is to study the residuals both graphically and quantitatively.

A stationary series $\{\varepsilon_i\}_{i \in \mathbb{N}}$ is white noise if and only if its spectral density $g$ is constant $\sigma^2/(2\pi)$, where $\sigma^2$ is the variance of the white noise. Therefore, testing the white noise assumption of $\varepsilon_i$ is equivalent to testing $g(\omega) = \sigma^2/(2\pi)$ for all frequencies $\omega \in [0, 2\pi]$. Let $I(\omega)$ be a spectral density estimate, say, periodogram. Fisher's test statistic is given by

$$T = \frac{\max_{1 \leq k \leq n} I(\omega_k)}{\sum_{k=1}^n I(\omega_k)/n},$$

where $\omega_k = 2\pi k/n$. A large value of $T$ indicates rejection of the null. Under the null hypothesis, normalized $T$ has an asymptotic extreme value type limiting distribution. See Section 7.4.1 in Fan and Yao [77].

Another test of white noise is based on the fact that a stationary white noise has zero autocorrelations for non-zero lags. Therefore, one can construct test statistics to measure discrepancies between sample autocorrelations and zero. Box and Pierce [43] and Ljung and Box [131] use the following test statistic:

$$T_m^{(1)} = n \sum_{k=1}^{m} \hat{\rho}(k)^2 \quad \text{and} \quad T_m^{(2)} = n(n+2) \sum_{k=1}^{m} \frac{\hat{\rho}(k)^2}{n-k},$$

where $\hat{\rho}(k)$ is the estimated autocorrelation at lag $k$ and $m$ is the maximum lag. Under the null hypothesis that $\{\varepsilon_i\}$ is a sequence of iid random variables, both test statistics have asymptotic chi-square distributions. Other residuals based tests include Fan and Li [80], and Hong and White [103].

## 7.4. Generalized likelihood ratio test

In an attempt to address nonparametric model validations problem, Fan *et al.* [79] developed generalized likelihood ratio test (GLRT). The basic idea behind their test is to compare profile likelihood under alternative to maximum likelihood under the null. Let $f$ be the function or vector of functions of interest, and $\eta$ the nuisance parameters. Denote by $\ell(f, \eta)$ the logarithm of the likelihood for a given dataset. Suppose that we are interested in testing the null $H_0: f = f_\theta, \theta \in \Theta$. The GLRT works as follows:

(a) For given $\eta$, estimate $f$ by $\hat{f}_\eta$ nonparametrically.
(b) Estimate $\eta$ by maximizing likelihood $\ell(\hat{f}_\eta, \eta)$.
(c) Compute the profile likelihood $\ell(\hat{f}_{\hat{\eta}}, \hat{\eta})$.
(d) Under $H_0$, estimate $(\theta, \eta)$ by the maximum likelihood estimator $(\hat{\theta}, \hat{\eta}_0) = \text{argmax}_{\theta, \eta} \ell(f_\theta, \eta)$.
(e) Compute the difference between the profile likelihood and the maximum likelihood,

$$T = \ell(\hat{f}_{\hat{\eta}}, \hat{\eta}) - \ell(f_{\hat{\theta}}, \hat{\eta}_0).$$

A significant positive value of $T$ indicates the rejection of $H_0$.

The cutoff value of $T$ can be obtained by establishing asymptotic distribution of $T$. For example, Fan *et al.* [79] consider varying-coefficient model and show that the so-called Wilks phenomenon holds. That is, the asymptotic distribution of $T$ under null hypothesis does not depend on nuisance parameters. The latter property can be used to determine the cutoff value via either the asymptotic distribution or Monte Carlo simulation for better accuracy.

As commented by Fan and Yao [[77], pp. 406], GLRT has been developed for independent data, but the idea can be extended to time series data; see

Chapter 9 in their book. In fact, one can easily apply the GLRT procedure to model validations problem for models in Sections 2 and 3 in conjunction with the nonparametric function estimation methods in Section 6.2 and the parameter estimation methods in Section 5.

### 7.5. Simultaneous confidence band

Let $\mu$ be a function of interest. Suppose our goal is to test the null hypothesis $H_0 : \mu = \mu_\theta$ for a parametric form $\mu_\theta$ with unknown parameter $\theta$. One way of achieving this goal is to construct simultaneous confidence band (SCB) for $\mu$. For $\alpha \in (0, 1)$ and a pair of functions $\ell_n(\cdot)$ and $u_n(\cdot)$ based on data, we say that $\ell_n(\cdot)$ and $u_n(\cdot)$ is a SCB for $\mu$ on a bounded interval $[T_1, T_2]$ with asymptotically correct nominal level $(1 - \alpha)$ if

$$\lim_{n \to \infty} \mathbb{P}\{l_n(x) \leq \mu(x) \leq u_n(x) \text{ for all } x \in [T_1, T_2]\} = 1 - \alpha. \quad (7.3)$$

A typical value of $\alpha$ is 5%. With the SCB, we can test the parametric hypothesis $H_0 : \mu = \mu_\theta$ based on the following procedure:

(a) Construct $(1 - \alpha)$ nonparametric SCB for $\mu$: $[\ell_n(\cdot), u_n(\cdot)]$;
(b) Under $H_0$, apply parametric methods to obtain an estimate $\hat{\theta}$ of $\theta$;
(c) Check whether $l_n(x) \leq \mu_{\hat{\theta}}(x) \leq u_n(x)$ holds for all $x \in [T_1, T_2]$, that is, whether the parametric estimate $\mu_{\hat{\theta}}$ is entirely contained with the SCB. If so, then we accept $H_0$ at level $\alpha$. Otherwise $H_0$ is rejected.

The first work on SCB construction appears in Bickel and Rosenblatt [38] for nonparametric kernel density of iid data. The idea is extended to mean regression and time trend function SCB construction by Johnston [113], Härdle [98], Knafl *et al.* [121] and Eubank and Speckman [71] for iid data and Wu and Zhao [164] for non-stationary time trend with time series errors. More recently, Zhao and Wu [173] have successfully applied the SCB based approach for model validations problem to the S&P 500 index. They find that an AR(1)-ARCH(1) model is an adequate fit for the S&P 500 index returns.

### 7.6. Density based test

The basic idea of density based tests is to measure distance between the nonparametric density estimate and parametric density estimate with large distance indicating inadequacy of the parametric model. We can use two different densities: marginal density and transition density.

Consider data $\{X_{i\Delta}\}_{0 \leq i \leq n}$ from model (2.7). Under $H_0 : (\mu, \sigma) = (\mu_\theta, \sigma_\theta)$, the theoretical marginal density is given by $f_\theta = f_{\mu_\theta, \sigma_\theta}$ in (6.1). Model validation procedure based on the marginal density works as follows:

(a) Under $H_0$, obtain estimate $\hat{\theta}$ of the parameter $\theta$ using parametric methods in Section 5.

(b) Obtain the parametric density estimate $\hat{f}_\theta = f_{\hat\theta} = f_{\mu_{\hat\theta},\sigma_{\hat\theta}}$.
(c) Construct nonparametric kernel density estimate $\hat{f}$ as in (6.2).
(d) Compute certain distance between the parametric density estimate $\hat{f}_\theta$ and the nonparametric density estimate. For example,

$$D_n = \frac{1}{n}\sum_{i=1}^n [\hat{f}(X_{i\Delta}) - \hat{f}_\theta(X_{i\Delta})]^2.$$

Reject $H_0$ if $D_n$ exceeds certain level. The critical value can be obtained by either studying asymptotic behavior of $D_n$ or through bootstrap methods.

This idea is applied by Aït-Sahalia [2] to daily data of 7-day Euro-dollar deposit rate during time period 1 June 1973 to 25 February 1995. He rejects all existing parametric models and proposes the new model (2.6). Clearly, this idea can be extended to deal with time series models by utilizing the convolution density estimate in Section 6.1. Recently, Zhao [170] studies model validation problem by constructing nonparametric simultaneous confidence band (see also Section 7.5) for marginal density and checking whether the implied parametric density estimate is entirely contained within the constructed band. Zhao [170] demonstrates that this density band based approach is widely applicable. For other contributions on density based test, see Gao and King [88] and Hong and Li [102].

Marginal density only captures part of the distributional properties of stochastic processes. For a Markov chain, another natural choice is the transition density. For model validations based on transition density, one can use the same procedure listed above by replacing the marginal density by the corresponding transition density. As in Section 6.1, one often needs to turn to the Euler approximation scheme (5.2) or the approximation in Aït-Sahalia [4]. Aït-Sahalia *et al.* [6] propose a transition density based test for continuous diffusion models and jump diffusion models. One could argue that a more efficient model validation procedure would incorporate information from both marginal transition densities. This would be a future research direction.

### 7.7. Other tests

We mention some other representative works. Anderson [11] introduces a goodness-of-fit test based on spectral density. Chen *et al.* [49] propose an empirical likelihood based test which is shown to be asymptotic equivalent to the nonparametric curve regression based test in Section 7.2. More references are collected in Chapters 3 and 5 of Gao [87] where semi-parametric specifications are also studied.

## 8. Tools for asymptotics

### 8.1. Mixing conditions

The Markovian property and mixing conditions for Markov chain play an important role in large sample theory in financial econometrics. A popular mixing coefficient is the $\rho$-mixing coefficient. Let $\{X_t\}_{t\geq 0}$ be a continuous-time stationary Markov process. For a random variable $X$ we write $\|Z\| = [\mathbb{E}(Z^2)]^{1/2}$ if the latter is finite.

Denote by $\mathcal{G}_t$ and $\mathcal{G}^t$ the sigma fields generated by $\{X_s\}_{s\leq t}$ and $\{X_s\}_{s\geq t}$, respectively. Let $\mathcal{L}^2$ be the set of square integrable random variables. Then the $\rho$-mixing coefficient of $\{X_t\}_{t\geq 0}$ is defined as

$$\rho_t = \sup_{s\geq 0} \rho(\mathcal{G}_s, \mathcal{G}^{s+t}),$$

with

$$\rho(\mathcal{G}, \mathcal{H}) = \sup\{|\mathrm{Corr}(G,H)| : G \odot \mathcal{G}, H \odot \mathcal{H}, G, H \in \mathcal{L}^2\}.$$

where $G(H) \odot \mathcal{G}(\mathcal{H})$ means that $G(H)$ is measurable with respect to $\mathcal{G}(\mathcal{H})$. For a random variable $X$, denote by $\sigma(X)$ the sigma field generated by $X$. Since $\{X_t\}_{t\geq 0}$ is a stationary Markov process, by Theorem 4.1 in Bradley (1986),

$$\rho_t = \rho(\sigma(X_0), \sigma(X_t)) = \sup_{g_1(X_0), g_2(X_t) \in \mathcal{L}^2} \{|\mathrm{Corr}[g_1(X_0), g_2(X_t)]|\}. \tag{8.1}$$

Let $\{\mathcal{J}_t\}_{t\geq 0}$ be a family of operators defined by $\mathcal{J}_t g(x) = \mathbb{E}[g(X_t)|X_0 = x]$ for $\|g(X_t)\|_2 < \infty$. Assuming that $\mathbb{E}[g_1(X_0)] = \mathbb{E}[g_2(X_0)] = 0$. By the Cauchy-Schwartz inequality,

$$|\mathrm{Cov}[g_1(X_0), g_2(X_t)]| = |\mathbb{E}\{g_1(X_0)\mathbb{E}[g_2(X_t)|X_0]\}| \leq \|g_1(X_0)\|\|\mathcal{J}_t g_2(X_0)\|,$$

which entails by (8.1)

$$\rho_t \leq \sup_{\mathbb{E}[g(X_0)]=0} \frac{\|\mathcal{J}_t g(X_0)\|}{\|g(X_0)\|}.$$

Notice that the stationarity and Markovian property of $\{X_t\}_{t\geq 0}$ imply that $\{\mathcal{J}_t\}_{t\geq 0}$ forms a semigroup in the sense that $\mathcal{J}_{s+t} = \mathcal{J}_s \mathcal{J}_t$. If there exists a fixed $t_0 > 0$ and constant $\lambda \in (0,1)$ such that $\|\mathcal{J}_{t_0} g(X_0)\|_2 \leq \lambda \|g(X_0)\|_2$ holds for all measurable function $g$ satisfying $\mathbb{E}[g(X_0)] = 0$, then we say that the operator $\mathcal{J}_{t_0}$ is a strong contraction. Under this condition, the family $\{\mathcal{J}_t\}_{t\geq 0}$ of operators is exponentially contracting. That is, $\|\mathcal{J}_t g(X_0)\| \leq \|g(X_0)\|_2 O(\lambda^{t/t_0})$ for all $\|g(X_0)\| < \infty, \mathbb{E}[g(X_0)] = 0$ and $t > 0$; see Banon [25]. Consequently, the process $\{X_t\}_{t\geq 0}$ has an exponentially decaying $\rho$-mixing coefficient and is called geometric $\rho$-mixing.

Sufficient conditions under which the operators $\{\mathcal{J}_t\}_{t\geq 0}$ possess the exponentially contracting property have been obtained in Banon [25], Hansen and Scheinkman [97], Genon-Catalot *et al.* [90] to new a few. Therefore, under such conditions, the discrete observations $\{X_i\}_{i\geq 0}$ is $\rho$-mixing with exponentially decaying $\rho$-mixing coefficient. Polynomial mixing condition is also available in Veretennilov [157]. For limit theorems under various mixing conditions, see the papers by Bradley [44], Hannan [94], Jones [114], Davydov [54], Dehling *et al.* [58].

### 8.2. Physical dependence measure

Wu [160] proposes the concept of *physical dependence measure*, a very powerful tool in studying nonlinear time series, for example, (3.1) or (3.3) and (3.4). To fix the idea, consider stationary process given by

$$X_i = G(\ldots, \varepsilon_{i-1}, \varepsilon_i), \quad i \in \mathbb{Z}, \quad \varepsilon_i : \text{iid}, \tag{8.2}$$

where $G$ is a measurable function such that $X_i$ is well defined. Examples of (8.2) include the popular ARMA, fractional ARIMA linear models, the nonlinear NARCH model (3.1), the random iterated functions $X_i = F(X_{i-1}, \varepsilon_i)$ for a random map $F(\cdot, \varepsilon_i)$ that maps $X_{i-1}$ to $X_i$ depending on the innovation $\varepsilon_i$, among others.

Let $(\varepsilon'_i)_{i\in\mathbb{Z}}$ be a iid copy of $(\varepsilon_i)_{i\in\mathbb{Z}}$. Define

$$X'_i = G(\ldots, \varepsilon_{-1}, \varepsilon'_0, \varepsilon_1, \ldots, \varepsilon_i).$$

Then $X'_i$ is a coupled version of $X_i$ with the innovation $\varepsilon_0$ therein being replaced by the iid copy $\varepsilon'_0$. Following Wu [160], define the coupling coefficient:

$$\theta_q(i) = \|X'_i - X_i\|_q, \quad \Theta_q(n) = \sum_{i=0}^{n} \theta_q(i), \quad q > 0, \tag{8.3}$$

where, for a random variable $Z$, $\|Z\|_q = [\mathbb{E}(|Z|^q)]^{1/q}$ if the latter is finite. By the construction of $X_i$ and $X'_i$, $\theta_i$ can be viewed as the contribution of $\varepsilon_0$ in predicting the future value $X_i$. Therefore, $\Theta_q(n)$ can be viewed as the cumulative contribution of $\varepsilon_0$ in predicting the whole future sequence $X_i, 0 \leq i \leq n$. Let $\Theta_q(\infty) = \lim_{n\to\infty} \Theta_q(n)$. If $\Theta_q(\infty) < \infty$, then we may interpret $(X_i)_{i\geq 0}$ as a process with short-range dependence.

Dedecker and Prieur [56] consider the following coupling coefficients,

$$\theta_q^*(i) = \|X_i^* - X_i\|_q, \quad \text{where} \quad X_i^* = G(\ldots, \varepsilon'_{-1}, \varepsilon'_0, \varepsilon_1, \ldots, \varepsilon_i). \tag{8.4}$$

The difference between the two coupling versions $X'_i$ and $X_i^*$ of $X_i$ is that the former replaces $\varepsilon_0$ with $\varepsilon'_0$ while the latter replaces $\varepsilon_j$ with $\varepsilon'_j$ for all $j \leq 0$. See [56] for more details. If $q \geq 1$, by the triangle inequality, we have $\theta_q^*(i) \leq \sum_{j=i}^{\infty} \theta_q(i)$ and $\theta_q(i) \leq \theta_q^*(i+1) + \theta_q^*(i)$. If $\theta_q(i) = O(\rho^i)$ for some $\rho \in (0,1)$, then the two coupling coefficients are equivalent. In many other cases, $\theta_q(i)$ is often smaller than $\theta_q^*(i)$. Consider, for example, the linear process $X_i = \sum_{j=1}^{\infty} a_j \varepsilon_{i-j}$. If $\mathbb{E}(|\varepsilon_0|^q) < \infty$ for some $q \geq 1$, then $\theta_q(i) = O(|a_i|)$ and $\theta_q^*(i) = O(\sum_{j=i}^{\infty} |a_j|)$.

Let $S_0 = 0$ and $S_{nt} = S_{\lfloor nt \rfloor} + (nt - \lfloor nt \rfloor)X_{\lceil nt \rceil}, 0 \le 0 \le 1$, be the partial sum process of $(X_i)_{i=1}^n$. The following result is useful in studying asymptotic behavior of nonlinear time series.

(i) [Wu [160], Dedecker and Merlevède [55]]. Assume $\Theta_2(\infty) < \infty$, then

$$\{S_{nt}/\sqrt{n}, 0 \le t \le 1\} \Rightarrow \{\sigma W_t, 0 \le t \le 1\}, \tag{8.5}$$

where $\sigma = \|\sum_{i=0}^\infty [\mathbb{E}(X_i|\ldots,\varepsilon_{-1},\varepsilon_0) - \mathbb{E}(X_i|\ldots,\varepsilon_{-2},\varepsilon_{-1})]\|_2 < \infty$, and $\{W_t\}_{0 \le t \le 1}$ is a standard Brownian motion.

(ii) [Wu [161]]. Further assume that $\sum_0^\infty i\theta_q(i) < \infty$ for some $2 < q \le 4$. Then on a possible richer probability space, there exists a Brownian motion $\{W_t\}_{t \ge 0}$ such that $S_{nt}$ can be uniformly approximated by $W_{nt}$ in the following sense:

$$\sup_{0 \le t \le 1} |S_{nt} - \sigma W_{nt}| = O_{\text{a.s.}}[n^{1/4}(\log n)^{1/2 + 1/q}(\log \log n)^{2/q}]. \tag{8.6}$$

The convergence (8.5) and the approximation (8.6) have different ranges of applicability. The approximation in (8.5) can often be used in studying asymptotic behavior of parametric methods, for example, maximum likelihood estimate, least-square estimates, M-estimation, generalized moment method. The approximation in (8.6) is particularly useful in nonparametric statistical inferences. For example, in nonparametric inference for time trend function, one needs to deal with quantities of the following form

$$V_n(t) = \sum_{i=1}^n \omega_n(i/n, t) X_i, \tag{8.7}$$

where $\omega_n(i/n, t), 1 \le i \le n$, are non-negative weights summing to one. Due to the dependence in $(X_i)_{1 \le i \le n}$ and the non-stationarity introduced by $\omega_n(i/n, t)$, it is usually difficult to study $V_n(t)$ directly. Write $\omega_i = \omega_n(i/n, t)$. Assume that (8.6) holds with $q = 4$. By the summation by parts formula,

$$\begin{aligned}
V_n(t) &= \sum_{i=1}^n \omega_i(S_i - S_{i-1}) = \sum_{i=2}^n (\omega_{i-1} - \omega_i)S_{i-1} + \omega_n S_n \\
&= \sum_{i=2}^n (\omega_{i-1} - \omega_i)\sigma W_{i-1} + \omega_n W_n + O_{\text{a.s.}}(n^{1/4}\log n)\Omega_n \\
&= \sum_{i=1}^n \omega_i \sigma(W_i - W_{i-1}) + O_{\text{a.s.}}(n^{1/4}\log n)\Omega_n,
\end{aligned}$$

where $\Omega_n = \sum_{i=2}^n |\omega_i - \omega_{i-1}| + \omega_n$. For many nonparametric estimates with bandwidth $b_n$, $\Omega_n = O[(nb_n)^{-1}]$. See Wu and Zhao [164].

Because $\theta_q(i)$ is directly related to the data-generating mechanism, it often has tractable bound. For example, consider the linear process $X_i = \sum_{j=1}^\infty a_j \varepsilon_{i-j}$, then $\theta_q(i) = O(|a_i|)$ provided that $\mathbb{E}(|\varepsilon_0^q|) < \infty$. For nonlinear models defined by the recursive equation

$$X_i = R_{\varepsilon_i}(X_{i-1}) \tag{8.8}$$

for a random function $R_{\varepsilon_i}$ depending on $\varepsilon_i$. Wu and Shao [163] obtain sufficient conditions under which: (i) (8.8) admits a unique stationary solution of the form (8.2); and (ii) the function $G$ satisfies the *geometric moment contraction* property $\theta_q(i) = O(\rho^i)$.

### 8.3. Martingale decomposition

The physical dependence measure in Section (8.2) becomes more powerful in nonparametric inferences when it is used in conjunction with martingale decomposition. Here we shall illustrate the idea using nonparametric kernel density estimate. Let $\{X_i\}_{1 \leq i \leq n}$ be a stationary Markov chain with transition density function $p(x|x')$ and invariance density $f(x)$. Consider the popular nonparametric kernel density estimate

$$\hat{f}(x) = \frac{1}{nb_n} \sum_{i=1}^{n} K_{b_n}(X_i - x), \quad \text{where} \quad K_{b_n}(u) = K(u/b_n). \tag{8.9}$$

Let $\mathcal{F}_i$ be the sigma filed generated by $X_j, j \leq i$. By the Markovian property,

$$\mathbb{E}[K_{b_n}(X_i - x)|\mathcal{F}_{i-1}] = b_n \int K(u) p(x - ub_n|X_{i-1}) du.$$

Let

$$I_n(x) = \sum_{i=1}^{n} \{p(x|X_{i-1}) - \mathbb{E}[p(x|X_{i-1})]\}.$$

Then we have the decomposition

$$\sum_{i=1}^{n} [K_{b_n}(X_i - x) - \mathbb{E}K_{b_n}(X_i - x)] = M_n(x) + b_n \int K(u) I_n(x - ub_n) du,$$

where

$$M_n(x) = \sum_{i=1}^{n} \{K_{b_n}(X_i - x) - \mathbb{E}[K_{b_n}(X_i - x)|\mathcal{F}_{i-1}]\}.$$

Notice that $M_n(x)$ is a martingale with respect to $\mathcal{F}_n$, and therefore various martingale results are applicable. Using the physical dependence measure in Section 8.2, one can show that, for certain short-range dependent processes, $b_n \int K(u) I_n(x - ub_n) du$ is of order $O_{\mathrm{p}}(\sqrt{n b_n})$ and negligible relative to the martingale part $M_n(x)$. Similar martingale decomposition techniques have been successfully applied to nonparametric inferences in Zhao and Wu [173] and Zhao [170].

## Acknowledgements

I would like to thank the associate editor Donald Richards for many valuable suggestions that greatly improved the presentation of this paper.